# Measurement of D/H and $^{13}$C/$^{12}$C Ratios in Methane Ice on Eris and Makemake: Evidence for Internal Activity

W.M. Grundy[1,2], I. Wong[3,4], C.R. Glein[5], S. Protopapa[5], B.J. Holler[6], J.C. Cook[7], J.A. Stansberry[6], A.H. Parker[8], J.I. Lunine[9], N. Pinilla-Alonso[10], A.C. de Souza Feliciano[10], R. Brunetto[11], J.P. Emery[2], and J. Licandro[12]

1. Lowell Observatory, Flagstaff Arizona.
2. Northern Arizona University, Flagstaff Arizona.
3. NASA Goddard Space Flight Center, Greenbelt MD.
4. American University, Washington DC.
5. Southwest Research Institute, San Antonio Texas.
6. Space Telescope Science Institute, Baltimore Maryland.
7. Pinhead Institute, Telluride Colorado.
8. SETI Institute, Mountain View California.
9. Cornell University, Ithaca New York.
10. Florida Space Institute, University of Central Florida, Orlando Florida.
11. Université Paris-Saclay, CNRS, Paris, France.
12. Instituto de Astrofísica de Canarias (IAC), La Laguna, Spain.

## Abstract

James Webb Space Telescope's NIRSpec infrared imaging spectrometer observed the outer solar system dwarf planets Eris and Makemake in reflected sunlight at wavelengths spanning 1 through 5 microns. Both objects have high albedo surfaces that are rich in methane ice, with a texture that permits long optical path lengths through the ice for solar photons. There is evidence for $N_2$ ice absorption around 4.2 µm on Eris, though not on Makemake. No CO ice absorption is seen at 4.67 µm on either body. For the first time, absorption bands of two heavy isotopologues of methane are observed at 2.615 µm ($^{13}$CH$_4$), 4.33 µm ($^{12}$CH$_3$D), and 4.57 µm ($^{12}$CH$_3$D). These bands enable us to measure D/H ratios of $(2.5 \pm 0.5) \times 10^{-4}$ and $(2.9 \pm 0.6) \times 10^{-4}$, along with $^{13}$C/$^{12}$C ratios of $0.012 \pm 0.002$ and $0.010 \pm 0.003$ in the surface methane ices of Eris and Makemake, respectively. The measured D/H ratios are much lower than that of presumably primordial methane in comet 67P/Churyumov-Gerasimenko, but they are similar to D/H ratios in water in many comets and larger outer solar system objects.  This similarity suggests that the hydrogen atoms in methane on Eris and Makemake originated from water, indicative of geochemical processes in past or even ongoing hot environments in their deep interiors.  The $^{13}$C/$^{12}$C ratios are consistent with commonly observed solar system values, suggesting no substantial enrichment in $^{13}$C as could happen if the methane currently on their surfaces was the residue of a much larger inventory that had mostly been lost to space.  Possible explanations include geologically recent outgassing from the interiors as well as processes that cycle the surface methane inventory to keep the uppermost surfaces refreshed.

## Introduction

The outer solar system is endowed with a large and diverse population of small icy planets that provide rich opportunities for comparative planetology.  Those that are massive and cold enough to retain volatile ices such as nitrogen and methane are an especially interesting clan (Schaller & Brown 2007; Johnson et al. 2015).  Changing seasonal patterns of insolation drive sublimation and bulk transport of gravitationally bound volatile ices (Trafton et al. 1998).  The seasonal redistribution of these ices segregates them from darker materials (e.g. radiolytic compounds, meteoric dust).  This keeps their albedo high and temperature low.  The phase changes between solid and vapor enable thermal energy, primarily from sunlight, to produce mechanical effects, sculpting their surfaces and creating diverse landforms (e.g., Moore et al. 2017; Stern et al. 2018; Grundy 2019).  Deep inside bodies that contain substantial inventories of rock, heat from decay of radionuclides may power production of volatiles (McKinnon et al. 2021).  Objects that still possess abundant methane at present include Triton, Pluto, Eris, and Makemake, in order of increasing methane absorption band depths observed in ground-based spectra (e.g. Cruikshank et al. 1993; Owen et al. 1993; Licandro et al. 2006a,b; Brown et al. 2005, 2007; Dumas et al. 2007; Tegler et al. 2007, 2008, 2010, 2012; Alvarez-Candal et al. 2011).

Eris and Makemake, the bodies with the strongest methane absorption bands, can provide us with novel comparisons with other transneptunian dwarf planets.  Eris is intermediate in mass between Triton and Pluto (Brown & Schaller 2007) but slightly smaller (diameter 2326 km) than Pluto (2377 km) and with a higher bulk density than both (Sicardy et al. 2011; Holler et al.

2021). The higher density implies a rock-rich composition, providing abundant internal heat from decay of long-lived radionuclides. Eris orbits much further from the Sun than Pluto, at a mean distance of 68 AU, but its high eccentricity brings it to 38 AU at perihelion and 98 AU at aphelion, resulting in nearly a factor of 7 difference in incident sunlight over the course of its orbit. Photometric observations show that Eris has a very high albedo (Mueller et al. 2019) and exhibits very little photometric variability as it spins on its axis (e.g., Carraro et al. 2006; Maris & Carraro 2008). The low lightcurve amplitude makes the length of Eris' day difficult to determine. A variety of tentative rotational periods have been reported (e.g., Duffard et al. 2008; Roe et al. 2008) but a combination of ground- and space-based observations has recently revealed it to be synchronous with its satellite Dysnomia's ~15.8 day orbital period (Bernstein et al. 2023; Szakáts et al. 2023), longer than the ~6 day diurnal cycles of Pluto and Triton. If it is assumed that Eris's spin axis is aligned with Dysnomia's orbit pole, this would imply a high obliquity of 78°, with important implications for Eris's seasons (e.g., Holler et al. 2021).

Makemake is a somewhat smaller body with equatorial and polar axes in the range between 1400 and 1500 km (Ortiz et al. 2012; Brown 2013), intermediate in size between Pluto and Charon. Makemake is similar to Eris in having its perihelion at 38 AU, though its aphelion is much closer than that of Eris, at only 53 AU, owing to its smaller semimajor axis and eccentricity. Makemake is also like Eris in having a very high albedo and low lightcurve variability. Various rotation periods have been reported, ranging from 8 to 23 hours (e.g., Ortiz et al. 2007; Heinze & deLahunta 2009; Hromakina et al. 2019). The discovery of a satellite (Parker et al. 2016) opens the prospect of determining an accurate mass and density for Makemake. Makemake, Eris, and Pluto all share similar linear polarization at low phase angles (Belskaya et al. 2012), presumably related to the light scattering properties of their seasonally mobile methane ice.

The high abundance of methane raises important questions about Eris and Makemake. First, what is the methane's source? Because comets generally contain methane, we might assume that methane was accreted from the solar nebula into pebbles/planetesimals at large heliocentric distances (e.g., Schaller & Brown 2007), but it is also possible that methane might have been produced in the interiors of Eris and Makemake, by analogy with geochemical processes that have been proposed for Pluto (McKinnon et al. 2021) and Titan (Glein 2015). Second, how do they maintain their high albedos? Methane ice exposed to energetic space radiation is rapidly processed into heavier hydrocarbons and ultimately dark, reddish tholin-like macro-molecules (e.g., Thompson et al. 1987; Stern et al. 1988). Inspired by the diverse landforms of Pluto and Triton, a variety of resurfacing scenarios can be imagined. Seasonal sublimation and condensation cycles could distill methane from its darker, more refractory radiation products, enabling the volatile methane to remain on top (Hofgartner et al. 2019). This mechanism could make them "bladed planets" by analogy to Pluto's methane-rich bladed terrain. If the surface methane deposit is thick enough, convective glacial overturn could continually refresh its surface, making Eris and Makemake more like "Sputnik planets" (Grundy and Umurhan 2017) reminiscent of Pluto's Sputnik Planitia (e.g., McKinnon et al. 2016).

Other volatile ices provide additional pieces to the puzzle of dwarf planet origin and evolution. Either the presence or apparent absence of certain species can be illuminating. For example, we might ask how much $N_2$ and CO ices are present, and what is their influence on resurfacing processes? Where might they have come from? The presence of $N_2$ had been inferred indirectly from its influence on the $CH_4$ bands seen on Eris and Makemake (e.g., Licandro et al. 2006a,b; Merlin et al. 2009; Tegler et al. 2008, 2010; Lorenzi et al. 2015), but it had not previously been directly detected. CO has not been detected on either body, though it is seen on Triton and Pluto and is abundant in comets. Heavier hydrocarbons can be produced via radiolysis or photolysis of methane (Bennett et al. 2006). $C_2H_4$ and $C_2H_6$ have been reported on Makemake (Brown et al. 2007, 2015) though not on Eris.

## Observations and data reduction

James Webb Space Telescope (JWST) observed Eris on 2022 August 30 as part of Cycle 1 Guaranteed Time Observations (GTO) Program 1191. At that time Eris was 95.1 AU from JWST and 95.8 AU from the Sun with a solar phase angle of 0.5°. The observations used the integral field unit (IFU) of the Near-Infrared Spectrograph (NIRSpec) instrument. The IFU has a field of view of 3"x3" and a spatial pixel scale of 0.1". Pairs of dithered exposures were obtained with three medium-resolution grating-filter combinations in sequence – G140M/F100LP, G235M/F170LP, and G395M/F290LP – to provide continuous wavelength coverage from 0.98 to 5.25 µm at a resolving power ($\lambda/\Delta\lambda$) of about 1000. The total exposure times in the three spectral settings were 613, 613, and 1196 seconds, respectively, yielding a combined exposure time of 2422 seconds. To optimize the detector noise performance, the NRSIRS2RAPID readout method was selected, which intersperses reference pixel reads with science pixel readouts (Moseley et al. 2010; Rauscher et al. 2012).

Makemake was observed on 2023 January 29 as part of Cycle 1 GTO Program 1254, when the target had observer-centric and heliocentric ranges of 52.2 and 52.7 AU, respectively, and a solar phase angle of 1.0°. The same gratings, dither pattern, and readout method were used for this observation, with per-grating exposure times of 521, 934, and 1780 seconds and a combined exposure time of 3035 seconds.

Data reduction and spectral extraction were carried out using a dedicated pipeline developed for NIRSpec IFU spectral observations. The core data processing functions were handled by Version 1.11.3 of the official JWST pipeline (Bushouse et al. 2022), with relevant calibration reference files supplied from context *jwst_1100.pmap* of the JWST Calibration Reference Data System (CRDS). The raw uncalibrated images (*uncal.fits* files) were first passed through Stage 1 of the JWST pipeline (*calwebb_detector1*), which converts the ramps of non-destructive detector readouts into 2D count rate images (*rate.fits* files) that are corrected for the bias level, dark current, non-linearity, and cosmic ray effects. While the IRS[2] readout greatly reduces read noise, there are residual systematic offsets in noise level across the detector, which manifest as column striping in the count rate images. To address this, a column-by-column read noise correction was applied by masking on-sky pixels to isolate the non-illuminated region, calculating the 200-pixel-wide moving median along each column, and subtracting the median array from the column pixel values. The noise-corrected *rate.fits* files were then processed through Stage 2 of the JWST pipeline (*calwebb_spec2*) to arrive at a set of 6 flat-fielded, distortion-corrected, wavelength-calibrated, and flux-calibrated IFU data cubes (*s3d.fits* files) – one for each dither position and grating. An IFU cube contains a stack of spatially rectified 2D image slices, each one corresponding to a single wavelength in the uniformly spaced wavelength solution. Figure 1 shows examples of the IFU cubes, collapsed along the wavelength axis, for both Eris and Makemake. The JWST pipeline outputs the IFU data cubes with units of MJy sr$^{-1}$. Prior to spectral extraction, the cubes were multiplied by the corresponding spatial pixel area contained in the headers to convert to irradiance units (MJy pixel$^{-1}$).

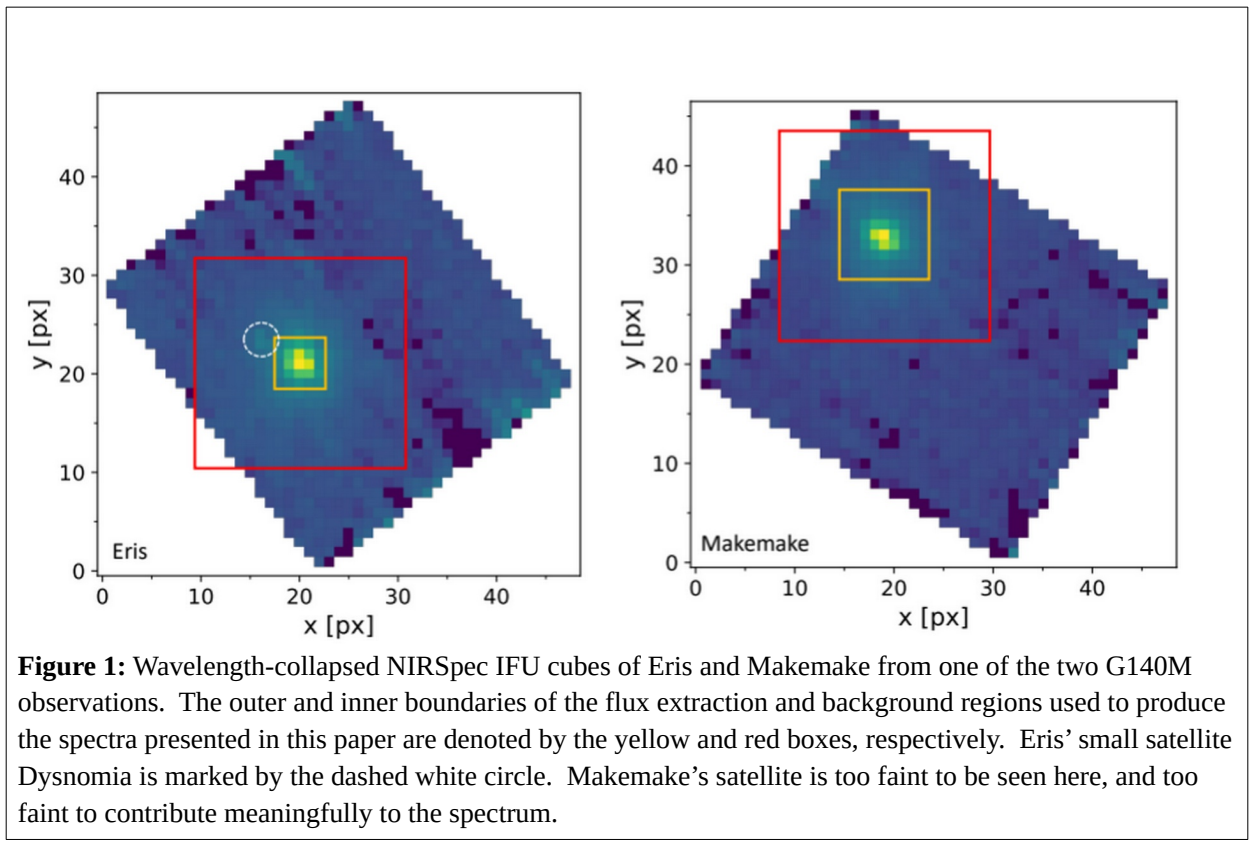

**Figure 1:** Wavelength-collapsed NIRSpec IFU cubes of Eris and Makemake from one of the two G140M observations. The outer and inner boundaries of the flux extraction and background regions used to produce the spectra presented in this paper are denoted by the yellow and red boxes, respectively. Eris' small satellite Dysnomia is marked by the dashed white circle. Makemake's satellite is too faint to be seen here, and too faint to contribute meaningfully to the spectrum.

Our data processing pipeline utilizes a specialized empirical point-spread function (PSF) fitting technique to extract the target's irradiance spectrum. The global source centroid was computed by collapsing the IFU cube along the wavelength axis and using a 2D Gaussian fit to estimate the peak position, rounded to the nearest integer pixel. To construct the template PSF at a given wavelength $k$, a median image slice was constructed by averaging image slices in the range [$k-W$, $k+W$], where $W$ is an adjustable window width. The background region was defined to be all pixels outside of a square box of side length $L_b$ centered on the centroid. All pixels with a nonzero data quality value were masked, and the median flux within the background region was subtracted from the median image. The template PSF of the target was defined within a box of side length $L_a$ centered on the centroid and was set equal to the flux of the median image slice, normalized to unit sum. All pixels within the outer background region were fixed to zero, while pixels between the regions defined by $L_a$ and $L_b$ were masked and not considered in the subsequent extraction. The source flux was then computed via a least-squares fit of the template PSF to the image slice $k$ using Levenberg-Marquardt $\chi^2$ minimization. The flux model consisted of the template PSF, multiplied by a scaling factor, along with an additive background level. Both the scaling factor and background level were free parameters in the fit, and the $\chi^2$ was weighted by the pixel flux uncertainties calculated by the JWST pipeline. The least-squares fitting was done iteratively, each time masking 5-σ outlier pixels until none remained. This PSF fitting procedure was carried out on every image slice, with the background-subtracted irradiance of the target at each wavelength provided by the best-fit scaling factor.

Our empirical PSF fitting technique is ideal for small body spectral extraction (i.e., point sources), because it provides greatly increased signal-to-noise over standard circular or square aperture extraction methods, while self-consistently accounting for wavelength-dependent changes in both source position and PSF shape and being more robust to bad pixels within the fitting region. We experimented with various values for $W$, $L_a$, and $L_b$. For NIRSpec IFU data cubes, there are wavelength-dependent oscillations in the source PSF width due to numerical artifacts that arise from the flux resampling during the cube building process. These oscillations can occur at scales as small as 30-40 image slices, so in order to accurately model the local PSF shape, we must use narrower window widths. We chose $W = 10$ when producing the spectra presented in this paper; varying the window widths between 5 and 15 produced no discernible systematic changes to the resultant flux values and overall scatter in the spectra.

When constructing template PSFs and extracting fluxes, we found that establishing a sizeable buffer region between the inner flux extraction region and the outer background region (i.e., $L_a < L_b$) yielded better signal-to-noise and reduced spectral scatter. As shown in Fig. 1, NIRSpec point-source PSFs have a large spatial extent. However, the outer regions of the diffraction pattern have relatively low signal-to-noise and are only several standard deviations above the overall background level. It follows that low-level detector artifacts and unflagged bad pixels can noticeably bias the template PSF scale factor in this region, which otherwise does not contribute appreciably to the total source flux. Moreover, in the case of Eris, the small satellite Dysnomia lies approximately 4 pixels from the primary centroid and can be discerned in the wavelength-collapsed IFU image slice (see Fig. 1). We therefore selected relatively compact flux extraction boxes. For Eris, $L_a = 5$ px was chosen for all dither positions and grating settings in order to avoid the peak of the secondary's PSF, while for Makemake, where the satellite is not discernible or expected to contribute significantly to the flux, we used $L_a = 9$ px. Varying $L_a$ by several pixels in either direction did not substantively affect the final spectra, even in the case of Eris, indicating that the contamination from Dysnomia's flux is negligible. In all cases, we set $L_b = 21$ px for the inner edge of the background region, which comfortably excludes all of the target PSF. The regions on the image slices corresponding to these spectral extraction parameter choices are indicated in Fig. 1. After extracting the irradiance spectrum for each dither position, we applied a 20 point wide moving-median filter to remove remaining 5-σ outliers and combined the two dither spectra together using a simple mean.

A small but non-negligible fraction of the total source PSF falls outside of the PSF fitting region denoted by $L_a$. We used NIRSpec flux calibration observations of the G2V-type standard star P330E obtained with the same grating settings (Cycle 1 Calibration Program 1538; PI: K. Gordon) to derive wavelength-dependent aperture correction curves and recover the full source irradiance of Eris and Makemake. After the uncalibrated data files from the standard star observation were reduced using an identical process to the Eris and Makemake data reductions, we extracted the star's flux using PSF fitting within boxes of sizes $L_a$ = 5 and 9 px. The resultant stellar spectra were then divided by the CALSPEC reference spectrum of P330E convolved to the sampling of the NIRSpec data (Bohlin et al. 2014) to compute the ratio array that quantifies the fraction of the total calibrated source flux within the corresponding extraction regions. By fitting cubic functions to these ratio arrays, we obtained empirical correction curves for each flux extraction region size and grating setting and divided them from the previously extracted spectra of Eris and Makemake to generate the corrected absolute flux spectra. We divided the irradiance of each object by the G-star spectrum that was extracted using the same $L_a$ value to get reflectance spectra. This process self-consistently accounts for both the flux losses outside of the extraction region and any minor wavelength-dependent instrumental systematics that are shared by all NIRSpec IFU spectra. A handful of visible outliers in the spectra remained at this stage, mostly attributable to poorly divided stellar lines, and we manually trimmed these. We scaled the reflectance to $I/F = \pi I r^2/F_\odot$, where $I$ is the radiance at the detector defined as power per unit area per unit solid angle, $r$ is the target heliocentric distance in AU, and $F_\odot$ is the solar flux at 1 AU. The latter was computed using the Planetary Spectrum Generator (Villanueva et al. 2018, 2022) and was scaled in a continuum region to the CALSPEC solar spectrum to recover units of MJy. $I$ was computed by diving the target flux corrected for light losses by the target area in pixel units and the pixel solid angle. The radii used to determine the target area of Eris and Makemake are 1163 and 715 km, respectively.

## Spectral interpretation and modeling

The complete JWST-NIRSpec reflectance spectrum of Eris is shown in Fig. 2. Previous visible and near-infrared reflectance spectra of Eris indicated a surface dominated by methane ice (e.g., Brown et al. 2005; Licandro et al. 2006b; Dumas et al. 2007; Abernathy et al. 2009; Merlin et al. 2009; Alvarez-Candal et al. 2011). These papers observed numerous methane vibrational combination and overtone bands between 1 and 2.5 µm that also appear in the JWST spectrum. JWST's extension of spectral coverage out beyond 5 µm reveals four strong infrared $CH_4$ ice bands that had not previously been detected at Eris. These bands, at 2.60, 3.32, 3.55, and 3.85 µm, correspond to the $3\nu_4$, $\nu_3$, $\nu_2 + \nu_4$, and $2\nu_4$ vibrational modes in methane ice (e.g., Ewing 1964; Calvani et al. 1989, 1992; Grundy et al. 2002). $CH_3D$ bands at 4.33 and 4.56 µm are also evident, corresponding to $2\nu_6$ and $\nu_2$ vibrational modes, respectively (Grundy et al. 2011). Much weaker $CH_3D$ bands at 2.47 ($2\nu_3 + \nu_5$) and 2.87 µm ($3\nu_6$ and $\nu_2 + \nu_3$) are marginally seen, too. Observation of these $CH_3D$ bands provides an opportunity to measure the D/H ratio in Eris' methane ice. Apart from $^{12}CH_4$, $CH_3D$, $^{13}CH_4$, and $N_2$ (see below), we have not unambiguously identified any other species in the new Eris spectrum. These non-detections include hydrocarbons like ethane, ethylene, and acetylene that are readily produced through radiolysis and photolysis of methane (e.g., Bennett et al. 2006), as well as $H_2O$, $CO_2$, $CH_3OH$, HCN, and $NH_3$ that have been detected on the surfaces of a number of other outer solar system bodies.

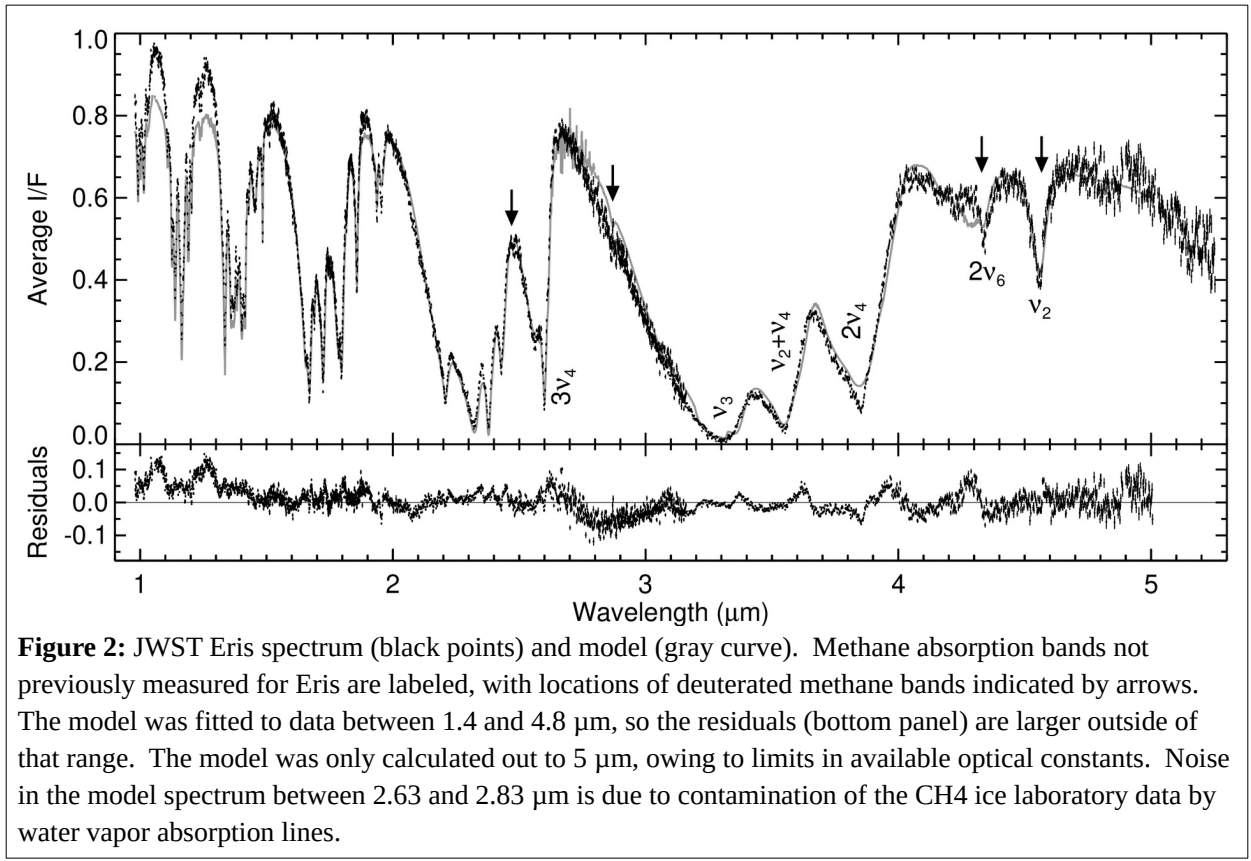

**Figure 2:** JWST Eris spectrum (black points) and model (gray curve). Methane absorption bands not previously measured for Eris are labeled, with locations of deuterated methane bands indicated by arrows. The model was fitted to data between 1.4 and 4.8 µm, so the residuals (bottom panel) are larger outside of that range. The model was only calculated out to 5 µm, owing to limits in available optical constants. Noise in the model spectrum between 2.63 and 2.83 µm is due to contamination of the CH4 ice laboratory data by water vapor absorption lines.

Quantitative information about solid planetary surfaces can be extracted from reflectance spectra by using a radiative transfer model to account for the wavelength dependent scattering of light by small particles in the planetary surface. We use the Hapke model for this purpose (e.g., Hapke 2012 and references therein). This model accounts for the wavelength-dependent optical behavior of individual regolith particles via two parameters: the single-scattering albedo ($w$) that expresses the fraction of light that is not absorbed in an interaction between a ray of light and a particle, and the single-scattering phase function ($P_{(g)}$) that describes the angular distribution of light scattering by the particle. We employ three commonly used simplifications to the model. First, we set the opposition effect and macroscopic roughness to fixed values, since these can generally only be constrained with multi-angular observations. Following Verbiscer et al. (2022) we set the opposition width parameter $h$ to 0.057, the amplitude parameter $B_0$ to 0.36, and the macroscopic roughness $\bar{\theta}$ to 20°. Second, we assume that only $w$ is a function of wavelength. Third, since we only need to evaluate $P_{(g)}$ at the single phase angle $g$ of the JWST observations, we treat $P_{(g)}$ as a constant rather than as a function of $g$, which is feasible in the IMSA (isotropic multiple-scattering approximation) variant of Hapke's model. To obtain $w$ for a particle of a specific material, we use the equivalent slab model described by Hapke (2012). Given a particle size and optical constants ($n$ and $k$) for a given wavelength, this model provides $w$ at that wavelength. An intimate mixture of different sorts of particles can be simulated by averaging their $w$ values, weighted by the particles' fractional contributions to the surface area.

Crucial model inputs are the optical constants of candidate materials at relevant wavelengths and temperatures. For methane ice we used the higher dynamic range absorption data from Grundy et al. (2002) along with the refractive index data from Gerakines & Hudson (2020). To treat the D/H ratio in the methane ice a free parameter we used the data and methodology of Grundy et al. (2011). This works by first removing the $CH_3D$ contribution present in the published methane ice optical constants, and then adding back in the appropriate amount of $CH_3D$ absorption corresponding to a specific D/H ratio, accounting for the fact that a methane molecule has four hydrogen atoms. Other optical constants used in this work include nitrogen ice, taken from a spectrum used in the analysis of Grundy et al. (2011) and carbon monoxide ice from Ehrenfreund et al. (1996).

The parameters of the model were iteratively modified to match the observed spectrum using the downhill simplex ("amoeba"; Nelder & Mead 1965) algorithm to minimize $\chi^2$ between model and observation, fitting to wavelengths from 1.4 to 4.8 µm, with wavelengths between 4.4 and 4.7 µm given triple weight to force the model to be especially attentive where the $\nu_2$ CH$_3$D band (4.56 µm) is relatively free of interference from other absorbers, making it the most diagnostic band for the D/H ratio. Where many bands of the same species are observed in a single spectrum, the particle size implied by the weaker bands tends to be larger than that implied by the stronger bands, indicating a more complex surface configuration than just a single particle size across the entire planet. Modelers employ a variety of strategies to handle this complexity, adding distinct terrain types, stratigraphic layers, or particle types. We used two different particle sizes for this purpose, which allowed the model to closely reproduce the data. The parameters that were adjusted were the D/H ratio in methane, the particle sizes and relative abundance of two sizes of methane ice particles, a scalar value for $P_{(g)}$, and nitrogen and carbon monoxide ice abundances. N$_2$ and CO were treated as minority contaminants dissolved in the methane ice by means of a volume-weighted linear mixture of the optical constants of pure methane plus the two contaminants.

The best fit Eris model is shown in Fig. 2. The derived parameters are D/H ratio $2.5 \times 10^{-4}$, methane grain sizes 0.39 and 0.02 cm diameter (with the smaller particles representing 3.2% of the total, by volume). Eris' high albedo requires a moderately high value for $P_{(g)}$ of 1.4. This could indicate a back-scattering single-scattering phase function, though it could also be a back-scattering lobe that is coupled with a forward-scattering lobe that we are not sensitive to owing to the small phase angle of the observation. Alternatively, the opposition effect could be stronger at these wavelengths in which case $B_0$ should be closer to unity. Any errors in the $I/F$ calibration would mostly be accommodated by shifts in the $P_{(g)}$ value. The model includes 22% N$_2$ ice by volume to account for a broad absorption between 4.0 and 4.3 µm, as shown in the left panel of Fig. 3. This absorption is weak, since N$_2$ is a homonuclear molecule without an intrinsic dipole moment. Interaction with neighboring molecules can induce a transient dipole, so this type of absorption band has been referred to as collisionally induced absorption (e.g., Van Kranendonk 1957; Shapiro and Gush 1966). There is no evidence for absorption by the vibrational fundamental band of CO ice at 4.67 µm. We should note that the available optical constants for CO and N$_2$ are not for those materials dissolved in CH$_4$ ice, but rather for the pure species. Studies of CO dissolved in N$_2$ ice do show modest differences in band shapes relative to pure CO ice (Quirico and Schmitt 1997a), so laboratory studies to investigate the spectral behaviors of both ices when diluted in $CH_4$ ice are called for.

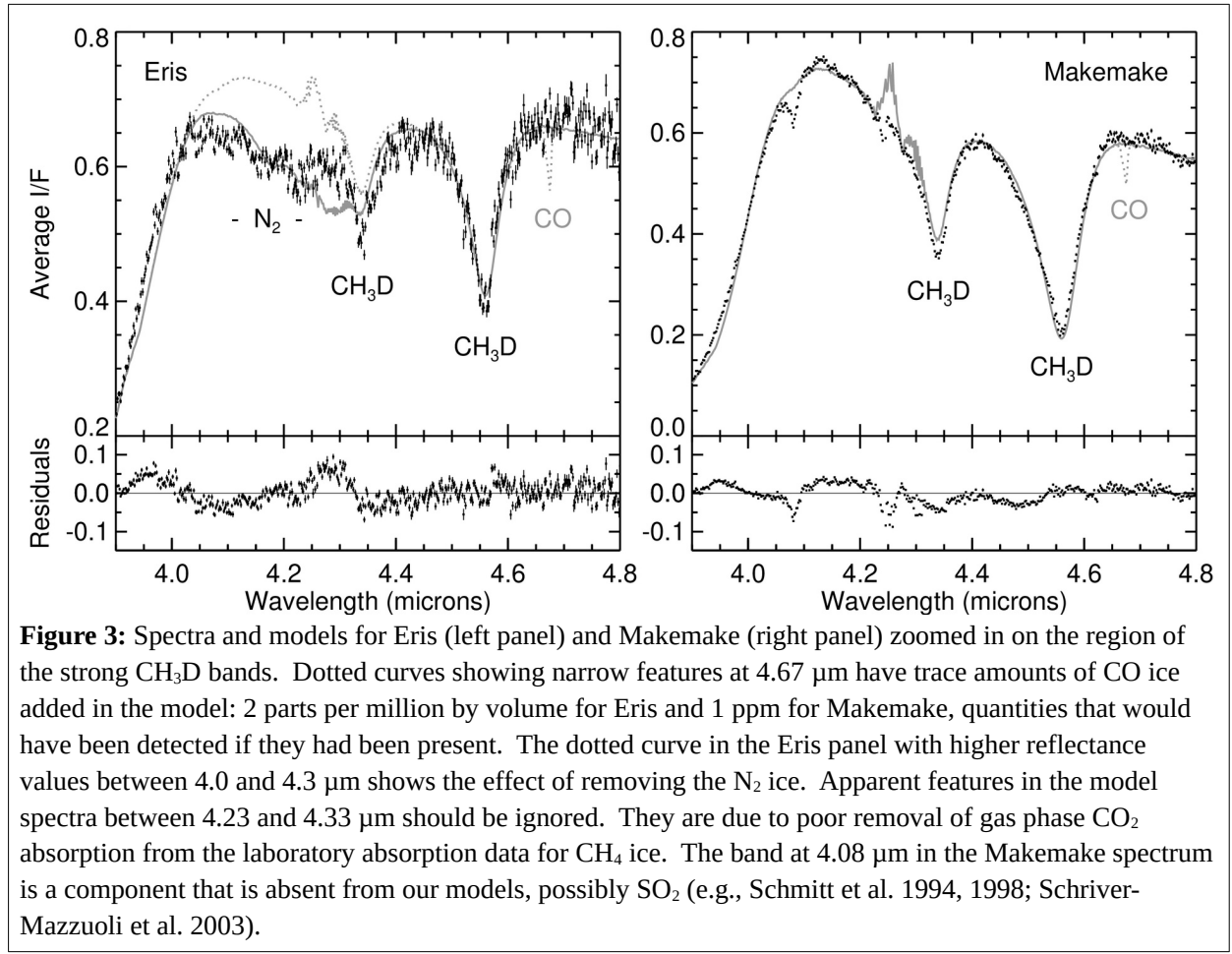

**Figure 3:** Spectra and models for Eris (left panel) and Makemake (right panel) zoomed in on the region of the strong $CH_3D$ bands. Dotted curves showing narrow features at 4.67 µm have trace amounts of CO ice added in the model: 2 parts per million by volume for Eris and 1 ppm for Makemake, quantities that would have been detected if they had been present. The dotted curve in the Eris panel with higher reflectance values between 4.0 and 4.3 µm shows the effect of removing the $N_2$ ice. Apparent features in the model spectra between 4.23 and 4.33 µm should be ignored. They are due to poor removal of gas phase $CO_2$ absorption from the laboratory absorption data for $CH_4$ ice. The band at 4.08 µm in the Makemake spectrum is a component that is absent from our models, possibly $SO_2$ (e.g., Schmitt et al. 1994, 1998; Schriver-Mazzuoli et al. 2003).

A portion of the Makemake spectrum was made available to us ahead of publication mainly for the purpose of isotopic analysis (S. Protopapa personal communication). Like Eris, the Makemake spectrum is methane-dominated and $CH_3D$ bands are clearly evident. We applied the same model to it, except with $B_0 = 1$, $h = 0.11$, and $\bar{\theta} = 5°$ (Verbiscer et al. 2022), and fitting to wavelengths between 3.4 and 4.8 µm (fitting to the Eris data in just this smaller range resulted in little change in D/H). The data and model are shown in the right panel of Fig. 3. Other materials not included in our model are clearly present in the Makemake spectrum, to be discussed in a follow-on paper, so the fit is not as clean as for Eris, and discrepancies are made more conspicuous thanks to the very high signal to noise ratio (S/N) of the Makemake data. For Makemake, we obtain best-fit parameters of D/H ratio $2.9 \times 10^{-4}$, methane grain sizes 1.2 and 0.06 cm (with the smaller particles representing 1.6% of the total, by volume), and $P_{(g)}$ equal to 1.8. The best fit model includes no $N_2$ or CO ice.

Uncertainties in the D/H ratios and other fitted parameters have various sources. Noise in the spectra is not the main source of uncertainty for D/H, since the spectra have relatively high signal-to-noise (S/N) where the $CH_3D$ bands appear. The main source of uncertainty comes from assumptions built into the radiative transfer model. It is necessary to choose a model configuration with some number of free parameters, as well as a wavelength range to fit the model to. Fitting to a larger wavelength range has the advantage of having more ordinary methane bands of varying depths to better constrain the texture of the methane ice, and thus its light scattering behavior that also applies to the $CH_3D$ dispersed in the $CH_4$. But that assumes no wavelength dependence of the scattering behavior such as occurs when there are textures at the scale of the wavelength (e.g., Hapke 2012), and no effects of different scattering in different regions of the surface, no vertical stratification, etc. To assess the uncertainty from model assumptions, we tested a variety of model configurations. These included models with wavelength- ($\lambda$) or $w$-dependent $P_{(g)}$, models using Hapke's internal scattering parameter $s$ set to a fitted constant multiplied by $\lambda^{-4}$ representing Rayleigh scattering by defects within particles, and models with three different particle sizes rather than two. Each model configuration was fitted to various ranges of wavelengths. We found that for all these different combinations the D/H ratio for each object varied within a band of approximately ±20%. The model presented in Fig. 2 is one that gave results near the middle of the range. We therefore report the D/H ratios for Eris and Makemake as $(2.5 \pm 0.5) \times 10^{-4}$ and $(2.9 \pm 0.6) \times 10^{-4}$, respectively. For $N_2$ abundance on Eris' surface, the range derived by the models was around 22 ± 5%, and for Makemake a few models returned abundances as high as 3%, so this could be taken as a 1-σ upper limit.

We see no evidence for CO ice absorption in either spectrum, which is somewhat surprising since it is readily seen in infrared spectra of Triton and Pluto (e.g., Cruikshank et al. 1993; Owen et al. 1993). The fundamental vibrational absorption band in pure α CO occurs at 4.67 μm. With the high S/N of the JWST spectra, and the absence of other spectral features at that wavelength, a volumetric upper limit of about 2 parts per million of CO can be placed for Eris, and less than 1 part per million for Makemake. But we reiterate the caveat that the spectral behavior of CO dissolved in $CH_4$ ice has not been studied in the laboratory. The vibrational band could be broadened or shifted in wavelength in that molecular environment, potentially changing these detection limits.

Small blue-shifts of a few Å in the $CH_4$ bands of Eris and Makemake have previously been reported (e.g., Licandro et al. 2006a,b; Abernathy et al. 2009; Alvarez-Candal et al. 2011; Lorenzi et al. 2015), and these shifts can be seen in the narrower methane bands in our JWST data, too. When $CH_4$ molecules are dispersed in $N_2$ ice, their vibrational absorption bands appear blue shifted (Quirico and Schmitt 1997b), but by a much larger amount than seen on Eris and Makemake. By fitting for the relative abundances of a shifted and an un-shifted $CH_4$ component, it is possible to estimate the fraction of $CH_4$ present in each phase. Where shifted and unshifted components are both seen and thermodynamic equilibrium is assumed, the solubility limits from the $N_2$-$CH_4$ binary phase diagram (Prokhvatilov and Yantsevich 1983) can be used to estimate the abundance of $N_2$ ice required to account for the fraction of methane seen in each component. Such an analysis was done for Eris (Tegler et al. 2010, 2012), implying a composition of 90% $N_2$ and 10% $CH_4$. That result is clearly at odds with the 22 ± 5% $N_2$ abundance we obtain from the JWST observation of the $N_2$ fundamental vibrational band for Eris. Likewise, the few Å shifts reported for Makemake's $CH_4$ bands attributed to the presence of $N_2$ could be seen as conflicting with our 3% upper limit for $N_2$ in Makemake's JWST spectrum. Several factors may account for these discrepancies, including the fact that small shifts in $CH_4$ bands have been found to occur when small amounts of $N_2$ are dissolved into $CH_4$ without exceeding the solubility limit so that no $N_2$ phase is present (Protopapa et al. 2015). Also, $CH_4$ bands can shift due to the presence of other impurities, such as argon, that lacks its own characteristic absorption bands and may or may not be present on Eris or Makemake (Tegler et al. 2010). Finally, as noted before, we lack laboratory optical constants for the $N_2$ absorption band when it is dissolved in $CH_4$, though data does exist for the $CH_4$ absorptions in such a mixture (Protopapa et al. 2015). Our ~22% abundance is based on using optical constants of pure β $N_2$ ice. This number exceeds the ~4% solubility limit (Prokhvatilov and Yantsevich 1983) of $N_2$ in $CH_4$ ice at Eris's ~30 K temperature, implying that an $N_2$ phase should also be present on Eris, presumably the lower temperature α phase. But relatively little $CH_4$ is soluble in α $N_2$ ice (approximately ~2%), so the presence of a small amount of this phase would have little effect on the appearance of the $CH_4$ bands. However, the shape of the $N_2$ fundamental band in α $N_2$ ice is somewhat different from that of β $N_2$, and is also temperature-dependent (Löwen et al. 1990; Schmitt et al. 1998). It can also be expected to be influenced by the presence of dissolved methane. More laboratory work is needed.

Replacing the carbon atom in methane with the heavier isotope $^{13}$C does not change the symmetry of the molecule as occurs in CH$_3$D, but the additional mass does shift the vibrational absorptions to slightly longer wavelengths. Since most methane bands are broad compared to the scale of the shift, a small fraction of $^{13}$CH$_4$ relative to ordinary methane is not easy to detect. What is required is an especially narrow band with a rapid transition to much weaker absorption at slightly longer wavelengths where the $^{13}$CH$_4$ version of the band occurs. The 2.60 µm CH$_4$ band seen for the first time for Eris and Makemake in the JWST spectra has just these characteristics. From preliminary laboratory experiments, this band is red-shifted by 0.015 µm in $^{13}$CH$_4$ (W. Grundy personal communication). We shifted the ordinary CH$_4$ optical constants by this amount to simulate optical constants of $^{13}$CH$_4$ and performed similar models as described earlier for CH$_3$D to find the best fit $^{13}$C/$^{12}$C ratios, using wavelengths between 2.57 and 2.65 µm. The best fit models are shown in Fig. 4, with $^{13}$C/$^{12}$C ratios of 0.012 ± 0.002 and 0.010 ± 0.003 for Eris and Makemake, respectively. These numbers are not statistically distinguishable from each other nor from the terrestrial inorganic standard VPDB (Vienna Pee Dee Belemnite) $^{13}$C/$^{12}$C ratio of 0.0112.

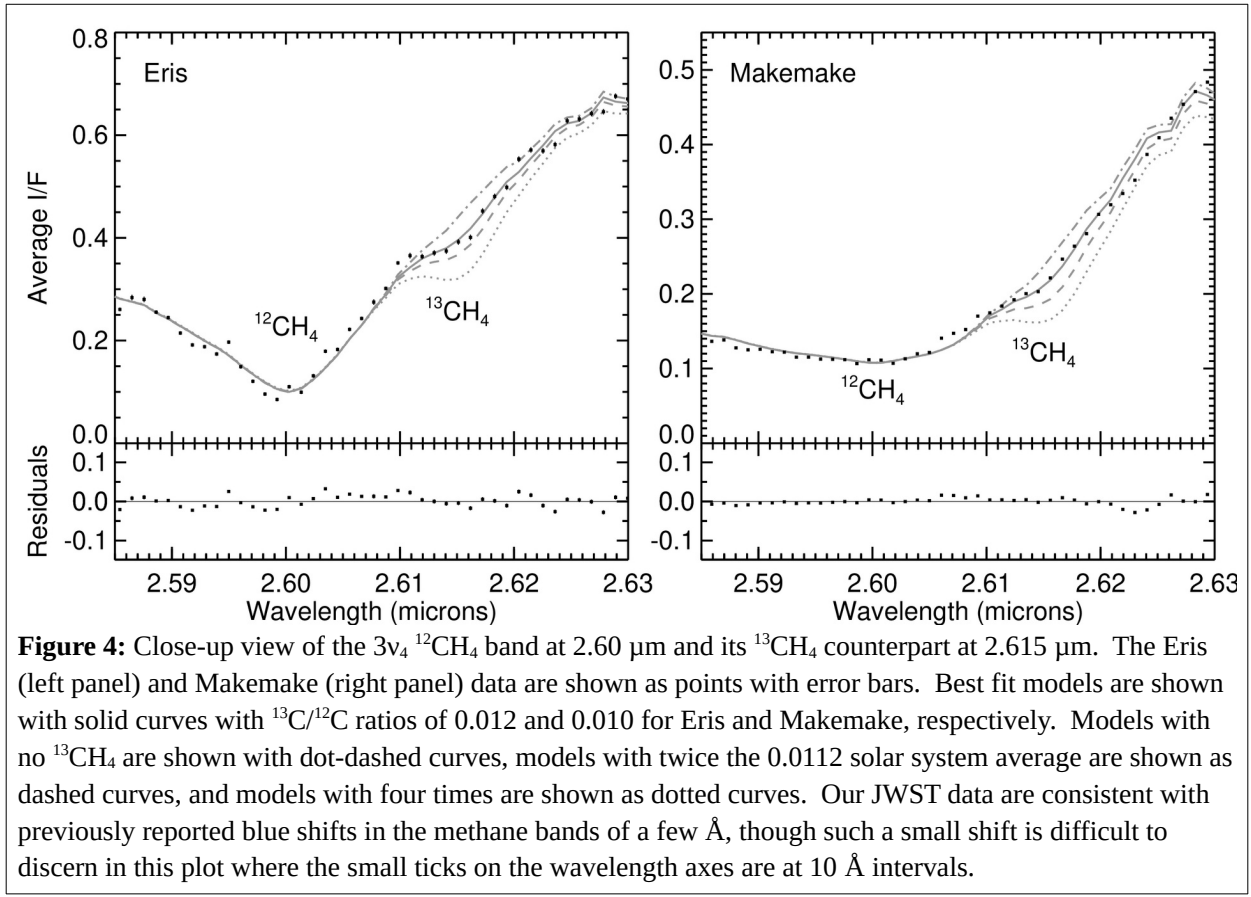

**Figure 4:** Close-up view of the 3ν$_4$ $^{12}$CH$_4$ band at 2.60 µm and its $^{13}$CH$_4$ counterpart at 2.615 µm. The Eris (left panel) and Makemake (right panel) data are shown as points with error bars. Best fit models are shown with solid curves with $^{13}$C/$^{12}$C ratios of 0.012 and 0.010 for Eris and Makemake, respectively. Models with no $^{13}$CH$_4$ are shown with dot-dashed curves, models with twice the 0.0112 solar system average are shown as dashed curves, and models with four times are shown as dotted curves. Our JWST data are consistent with previously reported blue shifts in the methane bands of a few Å, though such a small shift is difficult to discern in this plot where the small ticks on the wavelength axes are at 10 Å intervals.

Finally, we should remark on the startlingly large particle sizes for methane ice in our models, 0.4 and 1.2 cm for Eris and Makemake, respectively. Such enormous particles are unusual in a planetary regolith, and might be more properly thought of as a measure of the optical mean free path between defects in a polycrystalline, sintered slab of ice, rather than granular particles. The paucity of $N_2$ and CO may aid production of such large optical path lengths by not interfering with sintering and grain growth of $CH_4$, especially on Makemake (e.g., Eluszkiewicz et al. 2007).

## Geochemical interpretation and implications

We find D/H ratios in the methane ice of Eris and Makemake to be $(2.5 \pm 0.5) \times 10^{-4}$ and $(2.9 \pm 0.6) \times 10^{-4}$, respectively. These values can be compared with other reservoirs of hydrogen-bearing molecules in the solar system (see Fig. 5). Eris and Makemake values are somewhat higher than the terrestrial VSMOW (Vienna Standard Mean Ocean Water) value $(1.56 \times 10^{-4})$ and the D/H ratio of Titan's atmospheric methane $(1.59 \times 10^{-4}$, Nixon et al. 2012), and are much higher than the bulk protoplanetary nebula value that was dominated by hydrogen gas ($2.2 \times 10^{-5}$, e.g., Geiss and Gloeckler 2003; Aléon et al. 2022). But it is much lower than the $2.41 \times 10^{-3}$ D/H ratio of methane measured in the coma of comet 67P/Churyomov-Gerasimenko (Müller et al. 2022) or $1.57 \times 10^{-3}$ in the comet's refractory organic material (Paquette et al. 2021) not to mention isolated instances of extremely D-rich compositions in cometary and interplanetary dust particles (e.g., Aléon et al. 2001; McKeegan et al. 2006). Reduced carbon-bound hydrogen produced in the cold outer protosolar nebula or in interstellar environments generally appears to have high D/H ratios. This is a clue that methane seen on Eris and Makemake today was probably not delivered as a primordial species. Instead, their D/H ratios are similar to values determined for water and hydrated minerals in asteroids, comets, and larger icy bodies. This consistency suggests that hydrogen atoms in the methane ice that is now frozen on the surfaces of Eris and Makemake originally came from a source of water. A subsurface ocean of liquid water is one possibility (Hussmann et al. 2006). Another possible source of water-like D/H ratios are phyllosilicate minerals. These could have formed during a period of water-rock differentiation.

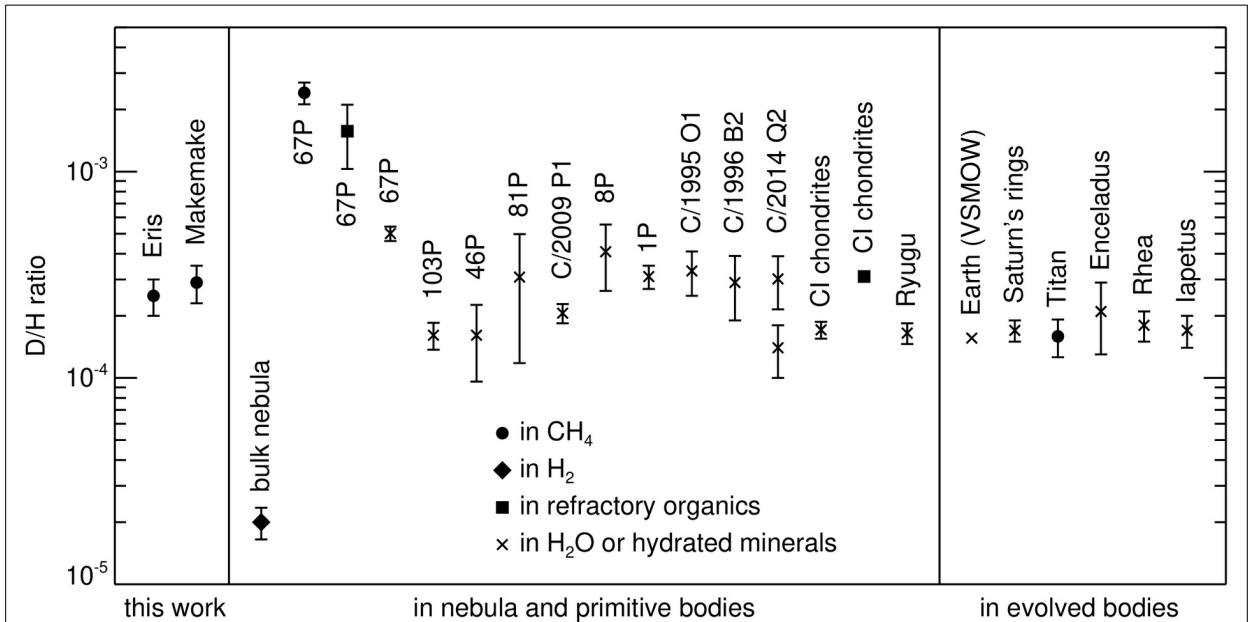

**Figure 5:** Eris and Makemake D/H ratios are greatly enriched compared with the bulk protoplanetary nebula (Geiss & Gloeckler 2003; Aléon et al. 2022), but much less enriched than methane or refractory organics in comet 67P/Churyumov-Gerasimenko (Paquette et al. 2021; Müller et al. 2022). They are more consistent with D/H ratios in water in comets (Bockelée-Morvan et al. 1998, 2015; Meier et al. 1998; Hartogh et al. 2011; Biver et al. 2016; Paganini et al. 2017; Lis et al. 2019). Hydrated minerals in CI chondrite meteorites and Ryugu (Piani et al. 2012, 2018, 2023) are a little lower, similar to ratios in $H_2O$ in Earth and the Saturn system (Waite Jr. et al. 2009; Clark et al. 2019), plus $CH_4$ in Titan's atmosphere (Nixon et al. 2012), while refractory organics in CI chondrites are more enriched (Alexander et al. 2007, 2012).

The clues afforded by our broad comparison in Fig. 5 can be placed on a firmer quantitative footing through modeling of D/H fractionation as described in the companion paper (Glein et al., submitted). Such modeling enables the D/H ratio of methane to provide a new window into the histories of Eris and Makemake. A more penetrating perspective of these previously dimly understood bodies can be achieved by determining which geochemical sources of methane are consistent with our derived D/H ratios. We highlight the main D/H modeling results in this first paper for the sake of continuity and convenience. More detailed information can be found in Glein et al. (submitted). We also discuss here key geochemical implications of the broader composition, including the $^{13}C/^{12}C$ ratio.

Glein et al. (submitted) conducted a detailed investigation to assess whether Eris' and Makemake's surface methane inventories could be primordial, abiotic, or thermogenic. Primordial methane means that these bodies would have acquired methane that was already in the form of methane (e.g., in accreted icy pebbles/planetesimals or via subsequent impact delivery). Abiotic methane (derived from $CO_2$ or CO) could be produced as a result of potential hydrothermal processes at the bottom of a subsurface ocean. Thermogenic methane would be produced by "cooking" accreted organic materials in a phyllosilicate-rich core that had undergone sufficient heating. If we can discriminate between these types of methane, then we can obtain useful information about formation conditions and processes that provide methane on icy dwarf planets.

Surprisingly, it was found that the simplest case of primordial methane is inconsistent with the data (Glein et al., submitted). If Eris' and Makemake's methane had a primordial origin, they would have high D/H ratios more like that of comet 67P (Müller et al. 2022). Instead, the observed D/H ratios point to a subsurface source that can generate abiotic or thermogenic methane. Since the production of both abiotic and thermogenic methane requires elevated temperatures to allow the necessary chemical reactions to occur, and because undifferentiated ice-rock mixtures at the relatively low pressures expected inside Eris ($\lesssim$15 kbar) and Makemake ($\lesssim$4 kbar) could not support such temperatures, we can conclude that their D/H ratios are evidence that these are differentiated planets with rocky cores that have been heated significantly (i.e., above a temperature of ~150°C; Stolper et al. 2014). It should be noted that we cannot give more specific preference to abiotic or thermogenic methane on the basis of present data, as both types of methane can provide consistent D/H ratios.

Other volatiles observed or not detected by JWST also indicate the importance of evolutionary processes in shaping the present volatile inventories of Eris and Makemake. Our detection of $N_2$ on Eris is consistent with $N_2$ formation from a $NH_3$ or an organic N source in the interior (Glein 2023). In seafloor hydrothermal fluids, $N_2$ production would be favored at higher temperatures and more oxidizing conditions, which could lead to $CH_4$ oxidation to $CO_2$ (Glein et al. 2008). We did not detect $CO_2$ on Eris, but there are conditions at which appreciable concentrations of $N_2$ and $CH_4$ can coexist at equilibrium (Glein et al., 2008). In addition, any $CO_2$ near the surface of Eris is likely covered by more volatile ices. Alternatively, Eris' $N_2$ could be associated with thermogenic processes in its inferred core. If conditions in the core were relatively hydrogen-poor, then the nitrogen speciation would be driven toward $N_2$ production (McKinnon et al. 2021).

On Makemake, JWST may not have detected $N_2$ owing to lower-temperature hydrothermal conditions (which would stabilize $NH_3$; Glein et al. 2008), as compared with Eris. Lower core temperatures would be consistent with the smaller size of Makemake. Another possibility is that Makemake might have previously had a $CH_4$-$N_2$ surface that evolved to a $CH_4$-rich surface by preferential escape of $N_2$. However, if Pluto is considered as an analogue, we might expect more $CH_4$ escape instead of $N_2$ escape (Young et al. 2018).

Hydrothermal geochemistry (e.g., Shock and McKinnon 1993) or early escape (e.g., Lunine and Nolan 1992) may explain a lack of CO on Eris and Makemake. Carbon monoxide is frequently the most abundant ice in comets after $H_2O$. Its extreme deficiency on Eris and Makemake would be hard to understand unless their present surface volatile compositions primarily reflect the role of evolution rather than primordial inheritance. The implication is that planet-sized Kuiper belt objects should not be regarded as "giant comets" that simply preserved their initial volatile inventories (cf. Glein and Waite 2018). Interior processes are called for to account for the D/H ratio in methane, and can also provide $N_2$ where it is present. Moreover, CO removal may be linked to $CH_4$ production if CO was a major carbon source of $CH_4$, as it is experimentally known to be (e.g., McCollom et al. 2010).

On the other hand, it may be more plausible to invoke early escape followed by volatile replacement. In this case, it could be assumed that the lightest and most volatile primordial species ($CH_4$, $N_2$, and CO) were all lost. Thus, high-D/H methane would no longer be present. Early rather than later, more continuous escape is hypothesized to get rid of CO on Eris, as present surface conditions (even at perihelion) are not conducive to CO escape (Schaller and Brown 2007). After early escape, the interior would have heated up until temperatures were sufficient to support endogenic production of $CH_4$ and $N_2$, which would have eventually been outgassed to the surface.

We determined carbon isotope ratios ($^{13}C/^{12}C$) of $0.012 \pm 0.002$ and $0.010 \pm 0.003$ for methane ice on Eris and Makemake, respectively. Carbon isotopes do not show much variability among solar system reservoirs. For example, methane in comet 67P/Churyumov-Gerasimenko has a $^{13}C/^{12}C$ ratio of $0.0114 \pm 0.0013$ (Müller et al. 2022), while the value for $CO_2$ is $0.0119 \pm 0.0006$ (Hässig et al. 2017). The protosolar value is thought to be ~0.0112 (Meibom et al. 2007), and organic matter in primitive carbonaceous chondrites has a $^{13}C/^{12}C$ ratio of ~0.0110 (Alexander et al. 2007). No matter how it originated, methane delivered to the surfaces of Eris and Makemake likely started with a $^{13}C/^{12}C$ ratio around 0.0112. The fact that the present inventories of methane have ratios that are indistinguishable from this value implies that no large $^{13}C$ enrichment occurred on either body.

Over time, atmospheric escape (and photo/radiation chemistry to a lesser extent; Nixon et al. 2012) will increase the $^{13}C/^{12}C$ ratio because of preferential loss of $^{12}CH_4$ owing to its smaller mass (and weaker carbon-hydrogen bond). Perhaps the simplest way to keep the isotopic ratio "normal" is for escape (and photochemical fractionation) processes to operate over a short duration. If methane was delivered to the surface geologically recently, then there would not be enough time for the isotope ratio to evolve to heavy values. Continuous delivery of methane would work similarly in shifting the ratio toward a reset. These scenarios are appealing as they could also explain the high albedos of Eris and Makemake, which would otherwise darken due to photochemically driven synthesis of complex organics. A deep convecting "Sputnik planet" layer of methane ice might also be able to account for these observations, by mixing heavier radiolytic products down into a much larger methane reservoir and preventing them from accumulating at the surface. If the $CH_4$ ice reservoir is sufficiently large, the same might be true for a "bladed planet" scenario with seasonal volatile transport cycles maintaining an uppermost visible surface dominated by lighter, more volatile species. With D/H ratios indicating an endogenic origin of methane (Glein et al., submitted), it is plausible to envision geologically recent outgassing processes delivering methane to the surface, analogous to previous proposals for Pluto (Howard et al. 2023) and Titan (Tobie et al. 2006). From this perspective, Eris would be more cryovolcanically active than Makemake, which would also be consistent with its less-red coloration and lack of detectable ethane, whereas Makemake is redder and shows evidence of irradiation products on its surface (Brown et al. 2007, 2015). On the other hand, the surface age of Makemake's methane may not be substantially older than that of Eris' methane, as there is no clear difference in their $^{13}C/^{12}C$ ratios.

Alternatively, Eris and Makemake may have relatively unevolved $^{13}C/^{12}C$ ratios because of slow orbit-integrated loss rates of methane, or their surface inventories of methane may be large so that loss processes have not made a noticeable dent in the isotopic ratio. Some combination of these factors and "recent" outgassing may be at work instead. Ultimately, understanding what recent, slow, and large mean for these worlds will require detailed evolutionary modeling to define limiting values, similar to work that has been done on Titan (i.e., Mandt et al., 2012; Nixon et al., 2012).

## Summary and conclusion

Our JWST infrared spectrum of Eris shows, for the first time, clear evidence of absorption by $CH_3D$ in methane ice at 4.33 and 4.56 μm, a subtle feature at 2.615 μm due to $^{13}CH_4$, and absorption by the fundamental band of $N_2$ ice between 4.0 and 4.3 µm. We also present the spectrum of Makemake in the region of the above $CH_3D$ and $^{13}CH_4$ bands. These data allow us to estimate the D/H and $^{13}C/^{12}C$ ratios in methane on Eris and Makemake, and constrain the abundance of $N_2$ ice on their surfaces. Unlike on Triton and Pluto, we find no evidence for CO ice on Eris or Makemake, with upper limits in the parts per million range.

Similar isotopic results are expected in the near future for methane ice on Pluto and Triton, and in water ice on Haumea and its family members, based on JWST spectra already in hand. These studies are providing our first look at isotopic ratios for the planet-sized bodies beyond Neptune with complex evolutionary histories and which appear to have undergone differentiation. Thanks to JWST, our knowledge of these bodies is undergoing dramatic advancement. More quantitative constraints on their histories will require future modeling of subsurface geophysical/geochemical conditions and the fate of primordial volatiles, as well as detailed comparisons with Pluto and Triton. Much smaller comets and (active) Centaurs can provide comparable data for much less evolved bodies, albeit based primarily on isotopic ratios in the gas phase.

It is interesting that we can learn something useful about the internal structures and physical conditions in the deep interiors of such remote worlds by measuring isotopic ratios. This would not have been possible without the JWST data reported in this paper. It has also not escaped our attention that some subsurface circumstances suggested by our results (e.g., hydrothermal activity) could support the generation of chemical disequilibria (Amend et al. 2011; Waite et al. 2017). There may now be an observational basis to consider Eris in particular as a possible ocean world and as the most distant candidate for habitability in the solar system.

## Acknowledgments

This work is based in part on observations made with the NASA/ESA/CSA James Webb Space Telescope. The data are at the MAST archive at STScI, which is operated by AURA, Inc., under NASA contract NAS 5-03127. These observations are associated with programs #1191 and #1254. We are especially grateful to Tony Roman, Shelly Meyett, and Charles Proffitt at STScI for help with program implementation. C.R.G. was supported by the NASA Astrobiology Institute through its JPL-led team entitled Habitability of Hydrocarbon Worlds: Titan and Beyond. J.A.S. acknowledges support through the sabbatical program at STScI, and the generosity of Lowell Observatory lodging in the Tombaugh apartment for 5 months, and NAU for office facilities during that time. N.P.A. acknowledges support from the Small Research Initiative of the Florida State operated by the Florida Space Institute. R.B. acknowledges support from the CNES. We also thank the free and open source software communities for empowering us with key tools used to complete this project, notably Linux, the GNU tools, LibreOffice, Evolution, Python, MariaDB, the Astronomy Users Library, and FVWM.

hydrothermal conditions. Geochim. Cosmochim. Acta 74, 2717-2740.

McKeegan, K.D., et al. (46 co-authors) 2006. Isotopic compositions of cometary matter returned by Stardust. Science 314, 1724-1728.

McKinnon, W.B., et al. (14 co-authors) 2016. Convection in a volatile nitrogen-ice-rich layer drives Pluto's geological vigour. Nature 534, 82-85.

McKinnon, W.B., C.R. Glein, T. Bertrand, and A.R. Rhoden 2021. Formation, composition, and history of the Pluto system: a post-New-Horizons synthesis. In: Stern, S.A., et al. (Eds.), *The Pluto System after New Horizons*. Univ. of Arizona Press, Tucson, AZ, pp. 507-543.

Meibom, A., A.N. Krot, F. Robert, S. Mostefaoui, S.S. Russell, M.I. Petaev, and M. Gounelle 2007. Nitrogen and carbon isotopic composition of the Sun inferred from a high-temperature solar nebular condensate. Astrophys. J. 656, L33.

Meier, R., T.C. Owen, H.E. Matthews, D.C. Jewitt, D. Bockelée-Morvan, N. Biver, J. Crovisier, and D. Gautier 1998. A determination of the $HDO/H_2O$ ratio in comet C/1995 O1 (Hale-Bopp). Science 279, 842-844.

Merlin, F., A. Alvarez-Candal, A. Delsanti, S. Fornasier, M.A. Barucci, F.E. DeMeo, C. de Bergh, A. Doressoundiram, E. Quirico, and B. Schmitt 2009. Stratification of methane ice on Eris' surface. Astron. J. 137, 315-328.

Moore, J.M., et al. (15 co-authors) 2017. Sublimation as a landform-shaping process on Pluto. Icarus 287, 320-333.

Moseley, S., R.G. Arendt, D. Fixsen, D. Lindler, M. Loose, and B.J. Rauscher 2010. Reducing the read noise of H2RG detector arrays: eliminating correlated noise with efficient use of reference signals. SPIE, 7742, 77421B.

Mueller, T., E. Lellouch, and S. Fornasier 2019. Trans-neptunian objects and Centaurs at thermal wavelengths. In: D. Prialnik, M.A. Barucci, L.A. Young (Eds.), *The Trans-Neptunian Solar System*, Elsevier, Cambridge MA, 153-181.

Müller, D.R., K. Altwegg, J.J. Berthelier, M. Combi, J. De Keyser, S.A. Fuselier, N. Hänni, B. Prestoni, M. Rubin, I.R.H.G. Schroeder I, and S.F. Wampfler 2022. High D/H ratios in water and alkanes in comet 67P/Churyumov-Gerasimenko measured with Rosetta/ROSINA DFMS. Astron. & Astrophys. 662, A69.

Nelder, J., and R. Mead 1965. A simplex method for function minimization. Computer Journal 7, 308-313.